\documentclass[usenatbib]{mn2e}
\input{epsf}
\usepackage{amssymb}
\usepackage{natbib}
\usepackage{color}
\usepackage{journals}

\newbox\grsign \setbox\grsign=\hbox{$>$} \newdimen\grdimen \grdimen=\ht\grsign
\newbox\simlessbox \newbox\simgreatbox
\setbox\simgreatbox=\hbox{\raise.5ex\hbox{$>$}\llap
     {\lower.5ex\hbox{$\sim$}}}\ht1=\grdimen\dp1=0pt
\setbox\simlessbox=\hbox{\raise.5ex\hbox{$<$}\llap
     {\lower.5ex\hbox{$\sim$}}}\ht2=\grdimen\dp2=0pt

\newcommand{\hMpc}{{\ifmmode{h^{-1}{\rm Mpc}}\else{$h^{-1}$Mpc }\fi}}
\newcommand{\hkpc}{{\ifmmode{h^{-1}{\rm kpc}}\else{$h^{-1}$kpc }\fi}}
\newcommand{\hMsun}{{\ifmmode{h^{-1}{\rm {M_{\odot}}}}\else{$h^{-1}{\rm{M_{\odot}}}$}\fi}}
\newcommand{\Msun}{{\ifmmode{{\rm {M_{\odot}}}}\else{${\rm{M_{\odot}}}$}\fi}}
\definecolor{grey}{rgb}{0.5,0.6,0.7}
\definecolor{purple}{rgb}{0.65,0.15,0.9}
\definecolor{darkorange}{rgb}{0.8,0.3,0}
\definecolor{olive}{rgb}{0.4,0.6,0.25}
\definecolor{darkgreen}{rgb}{0,0.7,0}

\title[Physical properties from satellite kinematics]{Physical properties 
underlying observed kinematics of satellite galaxies}
\author[R. Wojtak \& G. A. Mamon]{Rados{\l}aw Wojtak$^{1}$ and Gary A. Mamon$^{2}$
\\   \\
$^1$Dark Cosmology Centre, Niels Bohr Institute, University of Copenhagen, Juliane Maries Vej 30, DK-2100 Copenhagen \O, 
Denmark\\
$^2$Institut d'Astrophysique de Paris (UMR 7095: CNRS and UPMC),
    98 bis Bd Arago,
    F-75014 Paris, France \\
}

\begin{document}

\maketitle

\begin{abstract}
We study the kinematics of satellites around isolated galaxies selected from the Sloan Digital Sky Survey (SDSS) 
spectroscopic catalog. Using a model of the phase-space density
previously measured for the halos of $\Lambda$CDM dark matter cosmological simulations, we determine the properties of the halo mass 
distribution and the orbital anisotropy of the satellites as a function of the colour-based morphological type and 
the stellar mass of the central host galaxy. We place constraints on the halo mass and the concentration parameter 
of dark matter and the satellite number density profiles. We obtain a concentration-mass relation for galactic dark 
matter haloes that is consistent with predictions of a standard $\Lambda$CDM
cosmological model. At given halo or stellar mass, red galaxies have more concentrated halos than
their blue counterparts.
The fraction of dark matter within a few effective radii is minimal for $11.25
< \log M_\star < 11.5$. The number density profile 
of the satellites appears to be shallower than of dark matter, with the scale radius typically $60$ per cent larger 
than of dark matter. The orbital anisotropy around red hosts exhibits a mild excess of radial motions, in agreement 
with the typical anisotropy profiles found in cosmological simulations,
whereas blue galaxies are found to be
consistent with an isotropic velocity distribution. Our new constraints on
the halo masses of galaxies are used to provide analytic 
approximations of the halo-to-stellar mass relation for red and blue galaxies.

\end{abstract}


\section{Introduction}

According to the standard $\Lambda$CDM cosmological model, the majority of
the total energy density of the Universe is deposited 
in the form of dark energy and dark matter \citep{Komatsu+11,Rozo+10}. The former is a homogeneously distributed 
component responsible for the observed acceleration of the Universe expansion, whereas the latter is highly 
clumped, setting up a base for the growth of cosmic structures. Dark matter
(hereafter, DM)
assembles within quasi-spherical haloes that
host cosmological objects of all scale, from dwarf galaxies to clusters of galaxies. The properties of such 
haloes are one of the most fundamental predictions of the current
cosmological model. As first discovered by \cite*{NFW95,NFW96}, and confirmed
in many more recent and much better-resolved cosmological simulations
(e.g. \citealp{Springel+05,Springel+08,KTP11}), a key feature  
of DM haloes is the universal shape of their (hereafter NFW) density profile, whose logarithmic slope varies from 
$-1$ in the centre to $-3$ at large radii, while the transition scale between
these two slopes is correlated with 
the halo mass \citep{NFW97,Ludlow+11}. This property is the subject of various observational tests at all halo masses.

Studying the properties of DM distribution in galactic haloes is a challenge. Most methods rely on 
tracers whose positions coincide with the stellar component. Therefore, they probe only the inner 
part of an underlying gravitational potential of DM halo on the scale of a few per cent of the 
virial radius. Common means to study the inner part of DM density profiles is the measurement of 
rotation curves for spiral galaxies \citep{SR01}, the line-of-sight
velocity dispersion profiles of stars \citep{Bertin+94,Cappellari+06} or planetary
nebulae (e.g., \citealp{Napolitano+11}), 
strong \citep{Koopmans+06} and weak lensing \citep{Mandelbaum+06_profs,Gavazzi+07} or X-ray observations \citep{Humphrey+06} for early type galaxies. The main difficulty in interpretation 
of these data arises from the fact that the mass of DM is comparable to the baryonic component and, 
therefore, constraints on DM mass profile depends critically on the mass estimate of the stellar 
component \citep{ML05a}. In particular, the uncertainty of stellar population models, when applied to elliptical 
galaxies, leads unavoidably to an ambiguity about the shape of DM density
profile (e.g., \citealp{Grillo12}).

There are only two methods that allow to measure the DM distribution at distances comparable to the 
virial radius of the halo: weak lensing and kinematics of satellite galaxies.\footnote{Strong lensing is typically restricted to inner regions, while X-ray
measurements extend only to about half the virial radius.} Due to a very weak signal 
per galaxy, both methods rely on stacking the data, giving insight into the properties of a spherically averaged 
rather than individual haloes. 

Lensing analyses were successfully applied to measure projected mass profiles 
in galactic haloes. Most results are consistent with a universal NFW \citep*{NFW97} density profile of dark 
matter and the mass-concentration relation emerging from cosmological simulations of a standard $\Lambda$CDM 
model \citep{Mandelbaum+06_profs,MSH08}. Nevertheless, one weak lensing study
\citep{Gavazzi+07} concludes to a shallower DM density 
profile at large radii around massive elliptical galaxies, even slightly
shallower than the singular
isothermal sphere model with $\rho\propto r^{-2}$. Moreover, strong lensing
studies of ellipticals also point to a density profile with slope very close
to $-2$ between 0.3 and 0.9 $R_e$
\citep{Koopmans+06,Koopmans+09}.\footnote{We call $R_e$ the effective 
  radius, containing half of the luminosity in projection.} However, 
the na\"{\i}ve
superposition of NFW DM and the observed \cite{Sersic68} model for the stars,
leads to a slope 
close to $-2$ (from the model of \citealp{ML05b}, we predict a slope of
$-2.2\pm0.1$ in the range of radii studied by \citeauthor{Koopmans+06}):
the stars dominate the mass profile within $\approx 2\,R_e$ \citep{ML05b},
but at large radii the NFW DM component should dominate and the slope should
be considerably steeper than $-2$ ($\simeq -2.6$ at the virial radius from
the model above).

Satellite kinematics provide a popular means to estimate halo masses. Because
host galaxies possess a very small number of observable satellites, usually
one or two, one must stack the satellites over many host galaxies.
Still, early attempts \citep*{ZSFW93,ZW94} suffered from their small sample
sizes.
The Sloan Digital Sky Survey (SDSS) is the first very large
spectroscopic sample of galaxies with accurate redshifts and digital
photometry for a credible analysis of satellite kinematics.
 \cite{McKay+02} estimated host galaxy masses out to a fixed radius using
$M(r_{\rm ap}) = C r_{\rm ap}\sigma_{\rm ap}^2(r_{\rm ap})/G$, where $C$ is 
the logarithmic slope of the tracer density determined by fitting a power-law 
to the stacked satellite surface density profile, $r_{\rm p}$ is the aperture radius 
and $\sigma_{\rm ap}$ is the velocity dispersion inside the aperture. 
\cite{BS03} perform a similar analysis on the 2dFGRS, where they measure the
line-of-sight (LOS) velocity dispersion by fitting the LOS velocity
distribution by a Gaussian plus a constant term for interlopers (instead of
simply removing the high-velocity interlopers). They were the first to obtain
$M/L$ as a function of luminosity, and both for red and blue hosts. But their
analysis suffered from the relatively inaccurate velocities and photometry of
the 2dFGRS.
\cite{Prada+03} were the first to notice a decline of  LOS
velocity dispersion at projected radii. They 
showed that the distribution of satellites in projected phase
space (PPS) is consistent with the expectations from $\Lambda$CDM.
\cite{Conroy+07_dyn}  analysed the satellites from the Data Release 4 (DR4) of the SDSS and from the DEEP2 survey at
$z\approx 1$, again with a model for
the interlopers, and were the first to derive the variation of
virial mass with host galaxy luminosity, separating red and blue
galaxies. They found that red host galaxies of given blue luminosity have
double the halo mass as their blue counterparts.
\cite{More+11} added a second Gaussian to the Gaussian+flat distribution of LOS
velocities and fit aperture velocity dispersions (which are less sensitive
than LOS velocity dispersions to
the unknown orbital anisotropy and its radial variation) to find a halo
versus stellar mass in close agreement with that of \citeauthor{Conroy+07_dyn} \cite{Yeg11} showed 
that the halo mass from the velocity dispersion of satellites around spiral galaxies is consistent with that from the rotation curves extrapolated to large radii.

Unfortunately, all these analyses have flaws. For example, they all assume
that the LOS velocity distribution is a Gaussian (generally plus a uniform
interloper distribution), while it is known that anisotropic velocities lead
to non-Gaussian LOS velocities \citep{Merritt87}. Thus by taking into account
the non-Gaussian nature of the LOS velocity distribution \citep[see also][]{AmEv12}, one can both obtain
more accurate constraints on the mass profile and  derive constraints on the orbital anisotropy. 

We \citep{WLMG09} have recently developed a self-consistent method to derive at the same
time the mass and velocity anisotropy profiles of spherical systems. Our
method is based on the fact that the distribution of objects in PPS is a
triple integral \citep{DM92} over the LOS and the two plane-of-sky velocities of the
six-dimensional distribution function (DF) parameterised in terms of energy and 
angular momentum that \cite{Wojtak+08} measured on the halos of a $\Lambda$CDM simulation.
This approach gives much deeper insight into the data than tests of consistency shown before. 
It allows for a self-consistent comparison between a set of physical parameters determined from cosmological 
simulations and observations. Furthermore, analysis based on a PPS model does not rely on data binning which 
always introduces an artificial signal smoothing.

This manuscript is organised as follows. In section 2, we describe the data and criteria for selecting isolated 
galaxies and their satellites. Section 3 presents our dynamical model and a method of constraining parameters 
of the systems. The results of data analysis and discussion are presented in
section 4. The summary and a 
discussion follow in section 5. In this work, we adopted a flat $\Lambda$CDM cosmological model with 
$\Omega_{\rm m}=0.3$ and $H_{0}=70$ km s$^{-1}$ Mpc$^{-1}$.

\section{Data}
We made use of the Sloan Digital Sky Survey Data Release Seven (SDSS DR7, \citealp{Abazajian+09}) to select isolated 
galaxies and the satellite galaxies orbiting them. To search for the host galaxies, we considered a 
volume-limited subsample of the spectroscopic part of the survey defined by an $r$-band Petrosian absolute magnitude 
threshold $M_{r}=-19.0$ and redshift range of $3000 \textrm{ km s}^{-1}< cz< 25000\textrm{ km s}^{-1}$. 
The apparent magnitudes were converted to the 
absolute scale for our adopted cosmology and assuming colour-based $k$-corrections from \citet*{CMZ10}. 

We defined isolated 
central galaxies as those that are brighter by $\Delta M$ then every other galaxy lying inside an observational 
cylinder of a projected radius $\Delta R$ and a line-of-sight velocity range $2\Delta v_{\rm los}$. We fixed 
all fiducial parameters at values defining a rather restrictive criterion for galaxy isolation: $\Delta M>1.505$ 
(corresponding to the flux ratio of at least $2$), $\Delta R<1$ Mpc and $\Delta v_{\rm
  los}<1500$ km s$^{-1}$
(for comparison, see
\citealp{McKay+02,Prada+03,vdBNMY04,Conroy+07_dyn,KTP11}). 
All galaxies lying
in the cylinder 
and that are dimmer than the magnitude threshold are considered to be the satellites of the central galaxies. Due to 
a rather wide velocity cut-off, some of them are galaxies of background or foreground (interlopers). 
Disentangling between these two classes of galaxies is an intrinsic part of data analysis described in 
the following section.

We split the sample of the selected host galaxies into red and blue galaxies using $g-r+0.017M_{r}$ colour diagnostic 
(see \citealp{RBH10}), where $g$ and $r$ are $k$-corrected Petrosian magnitudes. A boundary value of this diagnostic 
was fixed at $0.25$ which is a minimum of the colour distribution lying between two Gaussian components corresponding 
to two galaxy populations. Using the publicly 
available catalog of the stellar mass estimates from the 
SDSS DR7,\footnote{http://www.mpa-garching.mpg.de/SDSS/DR7/Data/stellarmass.html} we found the masses of the stellar 
component of all central galaxies. Stellar masses were estimated using Bayesian approach as outlined in 
\cite{Kauffmann+03}, with fitting the observed photometry as described in \cite{Salim+07}. The model assumed 
initial mass function of \citet{Chabrier03}. We neglected $8$ per cent of
host galaxies for which stellar mass 
estimates were not available.

Our final sample consists of $8800$ and $2600$ satellites around $3800$ red and $1600$ blue
hosts, respectively. The host 
galaxies cover the stellar mass range from $\log_{10}(M_{\star}/\Msun)=10.0$ to $\log_{10}(M_{\star}/\Msun)=11.8$ 
for red galaxies and from $\log_{10}(M_{\star}/\Msun)=9.5$ to $\log_{10}(M_{\star}/\Msun)=11.0$ for the blue ones. 
Since the stellar mass is a better indicator of the haloes mass than galaxy luminosity \citep{More+11}, we split the 
sample of the host galaxies into several bins of the stellar mass. We used $6$ and $3$ bins for the red and blue 
hosts, respectively, as indicated in Table~\ref{tab-res}. This procedure
guarantees that the kinematic sample 
in every bin represents a homogeneous sample of DM haloes.

\section{Inference of physical parameters}
The velocity distribution of satellite galaxies is mostly determined by the gravitational potential of DM 
haloes of the central galaxy.
The second factor 
is the orbital anisotropy describing the fraction 
of radial-to-tangential orbits in the system. This additional degree of freedom makes data analysis more complex due 
to a well-known fact of the mass-anisotropy degeneracy \citep{BM82,MK90}. Breaking this degeneracy requires using 
rather complicated models accounting for higher-order corrections to the
Jeans equation (e.g., \citealp{MK90,Lokas02,LM03,WLMG09}). 
On the other hand, the advantage is that the same data allow to study two physical properties of the host-satellites 
systems at the same time -- mass distribution of DM halo and the orbital structure of the satellites 
(e.g., \citealp{LM03,Lokas09,WL10}).

\begin{figure}
\begin{center}
    \leavevmode
    \epsfxsize=8cm
    \epsfbox[50 50 580 420]{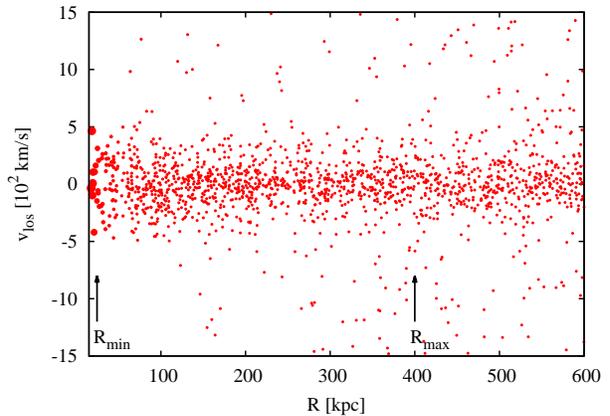}
\end{center}
\caption{Distribution of satellite galaxies in the projected phase space around isolated red galaxies 
with stellar masses $\log_{10}(M_{\star}/\Msun)=11.0-11.25$. The size of the symbols are inversely proportional 
to the completeness of the spectroscopic survey (proportional to the weights of data points in the analysis). 
The arrows show radius cut-offs adopted in the modelling.}
\label{diagram}
\end{figure}

We analysed the kinematic data in terms of the projected phase-space density, i.e. the density of satellite galaxies 
on the plane spanned by the line-of-sight velocity $v_{\rm los}$ and the projected distance from the host galaxy $R$ 
(see an example of the kinematical data in Fig.~\ref{diagram}). 
Due to projection effects, the $v_{\rm los}-R$ plane 
is populated by the true physical satellite galaxies of the central 
galaxies as well as interlopers. We did not apply any interloper removal to the data, but rather we accounted 
for the presence of interlopers in a statistical sense as an inherent part of a proper analysis.
This approach is particularly justified in case of the composite kinematic data for which the phase-space distribution 
of interlopers is smooth enough to be modelled by a continuous probability function, and is commonly adopted in many 
studies on kinematics of satellite galaxies \citep{Prada+03,Conroy+07_dyn,KP09}. The probability describing the phase-space distribution 
of interlopers introduces an additional degree of freedom to the proper model associated with the physical properties 
of the host-satellites systems. The observed projected phase-space density $p_{\rm los}(R,v_{\rm los})$ may be expressed 
as the following sum
\begin{equation}
\label{plos-gen}
g(R,v_{\rm los})=(1-p_{\rm i})\,g_{\rm sat}(R,v_{\rm los})+p_{\rm i}\,g_{\rm i}(R,v_{\rm los})\,,
\end{equation}
where $g_{\rm sat}(R,v_{\rm los})$ and $g_{\rm i}(R,v_{\rm los})$ are the projected phase-space densities 
of satellite galaxies and interlopers respectively, and $p_{\rm i}$ is the probability of a randomly picked 
galaxy being an interloper.

\subsection{Phase-space density model}
As a model of the phase-space distribution of satellite galaxies, we used an anisotropic model of the distribution 
function developed by \citep{Wojtak+08}. The model was designed to describe the phase-space properties of simulated 
DM haloes and it was successfully utilised to constrain mass profile and the orbital 
anisotropy in nearby galaxy clusters \citep{WL10}. The only modification required to adjust the model to the new context 
of systems of hosts and satellites is a non-constant ratio of DM-to-tracer density profile. Constant mass-to-light 
ratio appears to be a robust assumption in galaxy clusters \citep{BG03,LM03}, but it is not justified for the systems of 
satellite galaxies for which observations point to a bias between spatial distribution of DM and the 
satellite \citep{GCEF12}.

Following \citep{Wojtak+08}, we considered the phase-space density $f(\mathbf{r},\mathbf{v})$ of the following form
\begin{equation}
\label{DF}
f(\mathbf{r},\mathbf{v})=f_{E}(E)\,L^{-2\beta_{0}}\Big(1+\frac{L^{2}}{2L_{0}^{2}}\Big)^{(\beta_{0}-\beta_{\infty})}\,,
\end{equation}
where $E$ and $L$ are positively defined binding energy and angular momentum per unit mass, $\beta_{0}$ and $\beta_{\infty}$ 
are the asymptotic values of the anisotropy parameter at small and large radii, respectively. The anisotropy parameter 
quantifies the orbital anisotropy in terms of the ratio of the radial-to-tangential velocity dispersion and is traditionally 
defined as
\begin{equation}
\label{beta}
\beta(r)=1-\frac{\sigma^{2}_{\rm t}(r)}{2\,\sigma^{2}_{\rm r}},
\end{equation}
where $\sigma_{\rm r}$ and $\sigma_{\rm t}$ are the velocity dispersions in radial and tangential direction, respectively. 
The angular momentum part of the distribution function (\ref{DF}) is a generalisation of a well-know $L^{-2\beta}$ ansatz 
for systems with a constant anisotropy parameter \citep{Henon73_num}. It permits a wide family of the anisotropy profiles, suitable 
for constraining not only a global degree of the anisotropy, but also its radial profile. The anisotropy profiles are 
monotonic functions changing between two asymptotic values with the radius of transition given by $L_{0}$ parameter 
\citep[see][for details]{Wojtak+08}.

The energy part of the distribution function (\ref{DF}) is related to the distribution density of satellite galaxies 
$\rho_{\rm sat}(r)$ (the number density profile) and an underlying absolute value of the gravitational potential 
$\Psi(r)$ through the following integral equation
\begin{equation}
\label{DFE}
\rho_{\rm sat}(r)=\int\!\!\int\!\!\int f_{E}(E)\,
L^{-2\beta_{0}}\Big(1+\frac{L^{2}}{2L_{0}^{2}}\Big)^{(\beta_{0}-\beta_{\infty})}\textrm{d}^{3}v\,,
\end{equation}
where $E=\Psi(r)-\frac{1}{2}v^{2}$ is the binding energy. 
The mass profile at large distances from the central galaxies is dominated by DM, therefore we choose to neglect the contribution of stars (or in other words to assume that they are distributed like the DM). We checked that this 
assumption has negligible effect on our analysis (see Discussion).

We also approximated the DM density profile by the universal NFW profile \citep{NFW97} for which the 
gravitational potential takes the following form 
\citep{CL96,LM01}
\begin{equation}
\label{potential}
\Psi(r)=\Psi_{0}\frac{\ln \left(1+r/r_{\rm s}^{\rm DM}\right)}{r/r_{\rm s}^{\rm DM}}\,,
\end{equation}
where $r_{\rm s}^{\rm DM}$ is the scale radius at which logarithmic slope of DM density profile equals to $-2$. Our choice of 
the NFW parameterisation is motivated not only by cosmological simulations, but also by observational results showing consistency 
between satellite kinematics and dynamical predictions for DM haloes with the
NFW density profile
\citep{Prada+03,KTP11}. 
We note that the $\Psi_{0}$ and $r_{\rm s}^{\rm DM}$ are the principal parameters in the analysis which can be easily converted into 
more popular quantities describing the mass profile of DM haloes such as the virial mass or the concentration parameter 
(see \citealp{LM01} for all equations needed for parameter transformations).

The number density profile of satellite galaxies may be effectively approximated by the NFW profile with a scale radius 
unrelated to the concentration of DM (e.g., \citealp{GCEF12}). This property is independent of morphological type 
and the stellar mass of the host galaxy. Following this observational motivation, we adopted
\begin{equation}
\label{rhos}
\rho_{\rm sat}(r)\propto \frac{1}{(r/r_{\rm s}^{\rm sat})(1+r/r_{\rm s}^{\rm sat})^{2}}\,,
\end{equation}
as the number density of the satellites, where $r_{\rm s}^{\rm sat}$ is a new scale radius.

Having specified $\rho_{\rm sat}(r)$ and $\Psi(r)$ one can solve equation (\ref{DFE}) for the energy part of the distribution 
function. As shown by \cite{Wojtak+08}, the integral over velocity space may be reduced to a one-dimensional problem. Then, the 
resulting integral equation may be inverted numerically. We used the same scheme of the integral inversion as outlined in 
\citet{Wojtak+08}. Although the algorithm was designed to work for a single-component system with an NFW density profile, we 
checked that it is also feasible in case of two-component systems.

The projected phase-space density of satellite galaxies in (\ref{plos-gen}) was obtained by integrating the full phase-space 
density (\ref{DF}) over velocities $\mathbf{v_\perp}$ perpendicular to the line of sight and a spatial coordinate $z$ parallel 
to the line of sight \citep{DM92}
\begin{equation}
\label{plos-DF}
g(v_{\rm los},R)=2\pi R\int\textrm{d}z\int\!\!\int\textrm{d}^{2}v_\perp\,
f(E,L).
\end{equation}
Following the scheme outlined by \citet{WLMG09}, we calculated this integral numerically using Gaussian quadrature. We 
did not apply any fiducial truncation to the distance along the line of sight keeping the upper limit of the corresponding integral as 
defined by the condition of positive binding energy, i.e. $E=\Psi(\sqrt{R^{2}+z^{2}})-v_{\rm los}^{2}/2\geqslant0$.

\begin{figure}
\begin{center}
    \leavevmode
    \epsfxsize=8cm
    \epsfbox[50 50 580 420]{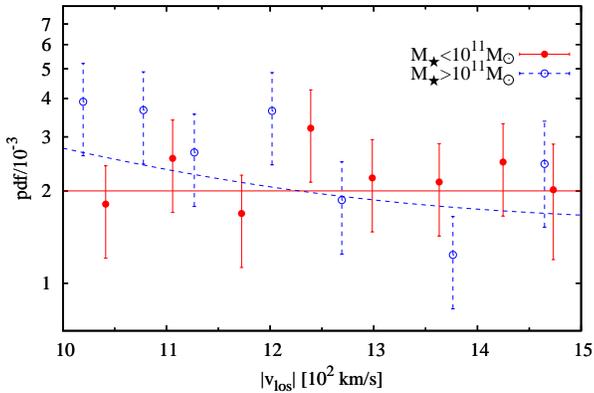}
\end{center}
\caption{Velocity distribution of the interlopers selected by positions $R>100$ kpc and velocities 
$|v_{\rm los}|>1000$ km~s$^{-1}$. Red (filled symbols, solid lines) and 
blue color (empty symbols, dashed lines) correspond to 
red hosts with stellar masses smaller and greater than $10^{11}\Msun$, respectively. The lines show 
a uniform model and a model with a broad Gaussian component (see eq.~\ref{plos-i}).}
\label{background}
\end{figure}

For interlopers, we adopted a uniform projected phase-space distribution, i.e. $g_{\rm i}\propto R$. It has been demonstrated that 
this model effectively separates gravitationally bound members of the central system from interlopers whose velocities are mostly 
dominated by the Hubble flow (\citealp{Wojtak+07}; \citealp*{MBM10}). 

As a consistency check, we verified the robustness of this model by comparing it with the 
distribution of galaxies at $R>100$ kpc and 
$1000\,\textrm{km s}^{-1}< |v_{\rm los}|<1500\,\rm km s^{-1}$, which constitutes a clear subsample of 
interlopers for all bins of the stellar mass ($|v_{\rm los}|$ exceeds maximum 
escape velocity at all halo masses as large as $10^{13.5}\Msun$). 
Although the surface density of these galaxies is consistent with a uniform distribution, the velocity distribution for massive red hosts ($\log_{10}(M_{\star}/\Msun)>11.0$) appears to exhibit a weak and broad non-uniform component 
(see Fig.~\ref{background}). 
We attribute this feature to  an environmental effect: massive red galaxies, although 
selected by the same isolation criteria, tend to populate denser environments which enhances the number of interlopers relative to low-density 
environments behind and in front of the host galaxy. Consequently, the interloper
density decreases with increasing velocity $|v_{\rm los}|$. 
We found that in order to account for this effect it suffices to consider the following modification of a uniform model
\begin{equation}
\label{plos-i}
g_{\rm i}\propto R\,\left\{\left(1-p_{\rm g}\right)\frac{1}{2v_{\rm max}}+p_{\rm g}\frac{1}{\sqrt{2\pi}\sigma_{\rm
      g}}\exp\left[-{v_{\rm los}^{2}\over 2\,\sigma_{\rm g}^{2}}\right] \right\}\,,
\end{equation}
where $p_{\rm g}$ is a nuisance parameter defining the relative weight of the Gaussian component, 
$2v_{\rm max}$ is the size of velocity cut-off ($3000$ km s$^{-1}$) and $\sigma_{\rm g}$ is the 
velocity dispersion induced by external gravitational field of the local environments. We used 
$\sigma_{\rm g}=500$ km s$^{-1}$, which is a typical velocity dispersion in the outskirts of galaxy clusters. 
Comparison between the model with the best-fit $p_{\rm g}$ from the final analysis ($p_{\rm g}\approx 0.7$) 
and the true velocity distribution of interlopers shown in Fig.~\ref{background} demonstrates that this 
form of background is sufficiently accurate to account for the observed effect of non-uniform velocity distribution of interlopers. 
This model was adopted for the analysis of the satellite kinematics in three bins of the most massive red host galaxies 
with $\log_{10}(M_{\star}/\Msun)>11.0$.
 
\subsection{Incompleteness}

Spectroscopic survey of the SDSS is not complete on angular scales smaller than $55''$ imposed by the minimum separation of 
the fibers in the spectrograph \citep{Blanton+03}. The limit of the completeness corresponds to the physical scale of $92$ kpc at 
the maximum redshift defining the sample of isolated galaxies. This distance is a substantial fraction of the virial radius and, 
therefore, it is relevant to correct the projected phase-space density model for the incompleteness. The correction was 
incorporated by means of weighting the projected phase-space density according to the local value of the completeness $w(R)$
\begin{equation}
\label{p-weighted}
\hat g\left(R,v_{\rm los}\right)=\left[g\left(R,v_{\rm los}\right)\right]^{1/w(R)}\,.
\end{equation}

\begin{figure}
\begin{center}
    \leavevmode
    \epsfxsize=8cm
    \epsfbox[50 50 580 420]{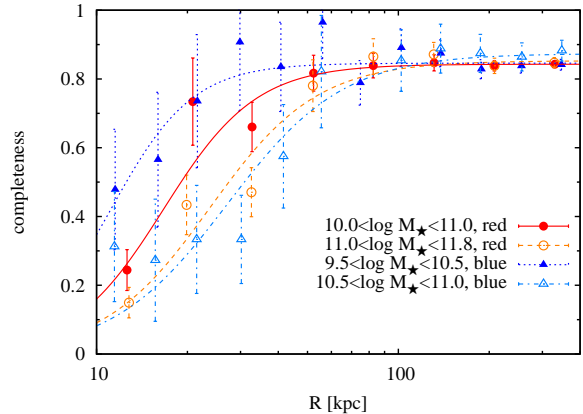}
\end{center}
\caption{Completeness of the SDSS redshift survey around isolated host
  galaxies. The data points are the measured values of completeness, obtained
  by comparing the counts of SDSS spectroscopic and photometric data. 
The colours (symbols and line types) represent different stellar mass bins of red and blues host 
galaxies as indicated in the legend. Solid lines are the best-fit
  profiles given by equation~(\ref{weight}) and Table~\ref{tab-wei}.}
\label{compl}
\end{figure}

\begin{table}
\begin{center}
\begin{tabular}{c|c|c|c}
$\log_{10}(M_{\star} [\Msun])$ & $R_{\rm w}\textrm{[kpc]}$ & $\alpha$ & $w_{0}$ \\
\hline
$10.0-11.0\textrm{(red)}$ & $16.7$ & $2.8$ & $0.84$ \\
$11.0-11.8\textrm{(red)}$ & $23.8$ & $2.5$ & $0.85$ \\
$9.5-10.5\textrm{(blue)}$ & $11.2$ & $3.2$ & $0.85$ \\
$10.5-11.0\textrm{(blue)}$  & $27.2$ & $2.3$ & $0.87$ \\
\end{tabular}
\caption{Best-fit parameters of the completeness given by equation (\ref{weight}).}
\label{tab-wei}
\end{center}
\end{table}

We measured the completeness of the data by computing the ratio of the surface number density of galaxies with spectroscopic redshifts 
to the surface density of all galaxies brighter than the magnitude limit of the SDSS spectroscopic survey, i.e. $r=17.7$ in $r$-band 
of Petrosian magnitude (see Fig.~\ref{compl}). In order to account for a redshift dependence we split the sample of red and blue 
galaxies into two bins of the stellar mass corresponding to two classes of the absolute magnitude. We found that the completeness 
may be well fitted by the following analytical form
\begin{equation}
\label{weight}
w(R)=w_{0}\frac{(R/R_{\rm w})^{\alpha}}{1+(R/R_{\rm w})^{\alpha}}\,,
\end{equation}
where $w_{0}$, $R_{\rm w}$ and $\alpha$ are free parameters. We fitted this
profile to the data in all mass bins of the host galaxies. 
The resulting best-fit profiles of $w(R)$ (see solid lines in Fig.~\ref{compl} and best-fit parameters in 
Table~\ref{tab-wei}) were used as the final weighting 
function in (\ref{p-weighted}). The final procedure of parameter inference was positively tested on incomplete mock data 
generated from the distribution function with a uniform background of interlopers.

\subsection{Parameter estimation}

We made use of kinematic data of satellite galaxies to place constraints on parameters of DM mass profile and the 
orbital anisotropy. 
For this, we adopted a Bayesian approach, maximising the likelihood of the
distribution of satellites in PPS.
The likelihood function $\mathcal{L}$ was defined as
\begin{eqnarray}
\nonumber
\ln\mathcal{L} & =  & \sum_{j=1}^{N}\ln\hat g \left(R_{j},v_{\rm los\;
  j}|\mathbf{a}\right) \,,\\
               & =  & \sum_{j=1}^{N}\frac{1}{w(R_{j})}\ln g\left(R_{j},v_{\rm los\; j}|\mathbf{a}\right)\,,
\end{eqnarray}
where the sum is over all satellite galaxies in a given stellar mass bin of the host galaxies and $\mathbf{a}$ is a vector 
of the model parameters. Both $g_{\rm sat}$ and $g_{\rm i}$ are normalised to $1$ over the area confined by velocity 
cut-off $|v_{\rm los}|<1500$ km s$^{-1}$ and radius range $[R_{\rm min},R_{\rm max}]$. As the maximum radius $R_{\rm max}$, 
we used the virial radius (see next paragraph) as it determines a natural boundary of the equilibrated part of DM 
haloes. Since the virial radius is a function of some model parameters, its value was estimated in an iterative approach starting 
with the best initial guess based on the halo-stellar-mass relation provided by \citet{Dutton+10}. For the most massive 
host galaxies, we imposed an additional limit $R_{\rm max}<400$ kpc which prevented from including the satellites which may be 
common to the host galaxy and local group or cluster of galaxies. We also
imposed a minimum radius, because 1) our correction to the spectroscopic
incompleteness is uncertain at small radii where our measured  completeness
is low, and 2) because photometric pipelines such as that of the SDSS tend
to fragment large (host) galaxies into one big one and many small ones
surrounding it, that appear like satellites, but are HII regions or spiral
arms instead. The minimum radius $R_{\rm min}$ was fixed at $5$ effective 
radii for red galaxies and $15$ kpc for the blue ones. Effective radii were estimated independently for every stellar mass bin 
using a scaling relation with the stellar mass found by \citet{HB09_curv}. The resulting minimum radius $R_{\rm min}$ changes 
from $10$ kpc for $\log_{10}(M_{\star}/\Msun)=10.0-10.5$ to $55$ kpc for $\log_{10}(M_{\star}/\Msun)=11.5-11.8$. 

\begin{table*}
\begin{center}
\begin{tabular}{c|c|c|c|c|c|c|c}
$\log_{10}(M_{\star} [\Msun])$ & $M_{r}\;[\textrm{mag}]$ & $\log_{10}(M_{100}[\Msun])$ & $c_{100}$ & $\log_{10}(M_{200}[\Msun])$ 
& $c_{200}$ & $r_{\rm s}^{\rm sat}/r_{\rm s}^{\rm DM}$ & $N_{\rm sat}/N_{\rm host}$ \\
\hline
$10.00-10.50(\textrm{red})$ & $-20.6$
                            &  $12.26_{-0.22}^{+ 0.03}$ &  $11.9_{ -7.4}^{+ 19.0}$ 
                            & $12.19_{-0.20}^{+ 0.03}$  &  $8.9_{ -5.8}^{+ 14.7}$ &  $1.7_{ -1.2}^{+  3.0}$ & $128/105$ \\
 & & & & & & & \\
$10.50-10.75(\textrm{red})$  & $-21.3$
                             & $12.40_{-0.08}^{+ 0.05}$ &  $16.7_{ -8.9}^{+  8.5}$ 
                             & $12.34_{-0.08}^{+ 0.04}$ &  $12.7_{ -6.9}^{+  6.6}$ &  $4.0_{ -1.6}^{+  3.7}$ & $303/242$ \\
 & & & & & & & \\
$10.75-11.00(\textrm{red})$  & $-21.8$
                             & $12.58_{-0.05}^{+ 0.02}$ &  $10.6_{ -3.6}^{+  5.5}$ 
                             & $12.51_{-0.05}^{+ 0.03}$ &  $7.9_{ -2.8}^{+  4.3}$ &   $3.4_{ -1.4}^{+  1.6}$ & $740/524$ \\
 & & & & & & & \\
$11.00-11.25(\textrm{red})$ & $-22.3$
                            &  $12.87_{-0.07}^{+ 0.03}$ &  $5.6_{ -2.4}^{+  5.3}$ 
                            &  $12.78_{-0.06}^{+ 0.03}$ &  $4.1_{ -1.8}^{+  4.1}$ &   $1.4_{ -0.6}^{+  1.3}$ & $1072/645$ \\
 & & & & & & & \\
$11.25-11.50(\textrm{red})$ & $-22.7$
                            & $13.19_{-0.10}^{+ 0.07}$ &   $2.6_{ -1.4}^{+  2.5}$ 
                            & $13.04_{-0.09}^{+ 0.10}$ &   $1.8_{ -1.1}^{+  1.9}$ &   $0.8_{ -0.4}^{+  0.9}$ & $726/357$ \\
 & & & & & & & \\
$11.50-11.80(\textrm{red})$  & $-23.2$
                             & $13.68_{-0.40}^{+ 0.06}$ &  $6.9_{ -5.6}^{+ 15.5}$ 
                             & $13.60_{-0.30}^{+ 0.09}$ &  $5.1_{ -4.2}^{+ 12.1}$ &   $1.5_{ -1.2}^{+  4.3}$ & $248/72$ \\
\hline
$9.50-10.00(\textrm{blue})$  & $-20.3$
                             & $11.82_{-0.43}^{+ 0.13}$ &  $16.1_{-15.0}^{+ 16.6}$ 
                             & $11.76_{-0.42}^{+ 0.10}$ &  $12.2_{-11.5}^{+ 12.9}$ &  $1.9_{ -1.7}^{+  2.3}$ & $44/42$ \\
 & & & & & & & \\
$10.00-10.50(\textrm{blue})$  & $-21.0$
                              & $12.07_{-0.13}^{+ 0.06}$  & $7.4_{ -4.3}^{+  4.5}$ 
                              & $11.99_{-0.11}^{+ 0.06}$  & $5.5_{ -3.3}^{+  3.5}$ &  $1.6_{ -1.0}^{+  1.4}$ & $159/142$ \\
 & & & & & & & \\
$10.50-11.00(\textrm{blue})$  & $-21.9$
                              & $12.41_{-0.12}^{+ 0.05}$ &  $4.5_{ -2.6}^{+  3.7}$ 
                              & $12.30_{-0.10}^{+ 0.06}$ &  $3.3_{ -2.0}^{+  2.8}$ &  $1.3_{ -0.7}^{+  1.4}$ & $323/263$ \\

\end{tabular}
\caption{Constraints on the parameters of the DM density profile and the number density of galaxy satellites in all bins 
of stellar mass $M_{\star}$ and both types of host galaxies (red and blue): the range of stellar mass $M_{\star}$, 
the median $r$-band absolute magnitude, the DM halo mass $M_{100}$ (or $M_{200}$), the concentration 
parameter $c_{100}$ (or $c_{200}$) and the ratio of tracer-to-DM scale radius 
$r_{\rm s}^{\rm sat}/r_{\rm s}^{\rm DM}$. The table 
provides the best fit values at the maximum of the posterior probability and the ranges containing $68$ per cent of the 
corresponding marginal probability. The last column contains the number of satellites between $R_{\rm min}$ and 
$R_{\rm max}$, $N_{\rm sat}$, and the number of hosts, $N_{\rm host}$.}
\label{tab-res}
\end{center}
\end{table*}

Analysis of the likelihood was carried out using the Markov Chain
Monte Carlo (MCMC) technique with the Metropolis-Hastings algorithm
\citep[see e.g.]{GCSR04}. The set of the primary parameters used in the MCMC analysis comprises the scale radius $r_{\rm s}^{\rm sat}$ of the number density of 
satellites (\ref{rhos}), the dimensionless $r_{\rm s}^{\rm sat}/r_{\rm
  s}^{\rm DM}$ ratio, the normalisation $\Psi_{0}$ of 
the gravitational potential (\ref{potential}), the asymptotic velocity anisotropy at small and large radii ($\beta_{0}$ and $\beta_{\infty}$, 
respectively), the interloper probability $p_{\rm i}$ in (\ref{plos-gen}) and
the relative weight $p_{\rm g}$ of the Gaussian part in the velocity 
distribution (\ref{plos-i}) of interlopers (applied only to the data of red galaxies with $M_{\star}>10^{11}\Msun$). 
We determined constraints on the mass profiles by converting parameters of the gravitational potential, i.e. $\Psi_{0}$ 
and $r_{\rm s}^{\rm DM}$, into the standard parameters characterising DM halo with the universal NFW density profile: 
the virial mass $M_{\Delta}$ and the concentration parameter $c_{\Delta}$. The virial mass is defined in terms of the mean density inside 
the sphere of radius $r_{\Delta}$ (the so-called virial radius) relative to the critical density $\rho_{\rm c}$
\begin{equation}
\frac{3M_{\Delta}}{4\pi r_{\Delta}^{3}}=\Delta \rho_{\rm c},
\end{equation}
where $\Delta$ is the virial overdensity. The concentration parameter is the virial radius expressed in the unit of the scale 
radius $r_{\rm s}^{\rm DM}$, i.e. $c_{\Delta}=r_{\Delta}/r_{\rm s}^{\rm DM}$. We adopted two commonly used values of the overdensity 
parameter: $\Delta=200$ and $\Delta=100$. The latter corresponds, to a $3$ per cent precision, to the virial overdensity of 
a standard $\Lambda$CDM cosmological model \citep{BN98}.

We carried out the MCMC analysis assuming log-uniform priors for $r_{\rm s}^{\rm sat}$, $\Psi_{0}$ and $r_{\rm s}^{\rm sat}/r_{\rm s}^{\rm DM}$, 
and uniform priors for all remaining parameters.  We fixed $L_{0}$ (eq.~[\ref{DF}]) at $0.2\sqrt{\Psi_{0}}r_{\rm s}^{\rm sat}$ 
corresponding to the $\sim 1\,r_{\rm s}^{\rm sat}$ transition radius between two asymptotic values of the anisotropy parameter 
\citep{WL10}. The central anisotropy $\beta_{0}$ was limited by $\beta_{0}<1/2$ in order to the prevent distribution function 
from taking negative values \citep{AE06_ApJ}. Constraints on all parameters are based on Markov chains containing $2\times 10^{4}$ 
models in every bin of the stellar mass. Every chain was preceded by a number
of trial chains ran to estimate the covariance matrix 
of the proposal probability distribution \citep{GCSR04}.

\section{Results}

\begin{figure}
\begin{center}
    \leavevmode
    \epsfxsize=8cm
    \epsfbox[50 50 580 420]{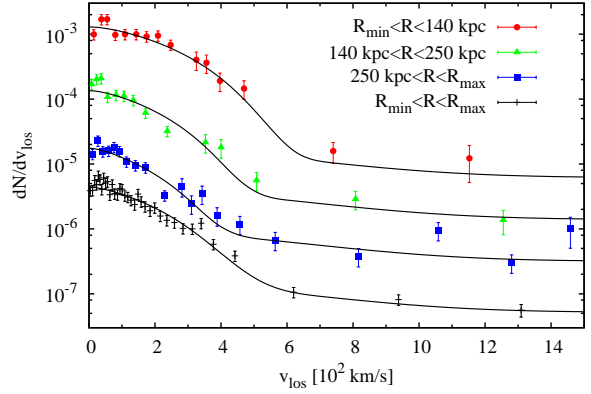}
\end{center}
\caption{Example of velocity distributions of the satellites around isolated red galaxies with 
stellar masses $\log_{10}(M_{\star}/\Msun)=11.0-11.25$. Black solid lines are the profiles 
calculated using the best-fit phase-space density model. Cut-off radii adopted in the analysis 
are $R_{\rm min}=24.5$ kpc and $R_{\rm max}=400$ kpc.}
\label{vdf}
\end{figure}

Table~\ref{tab-res} shows our constraints on the halo mass, the concentration parameter and the ratio of the tracer-to-dark-matter 
scale radius of the density profile, for all bins of the host
galaxies. Fig.~\ref{vdf} illustrates a goodness-of-fit test of our
  model: it compares the velocity distributions of the satellites around red hosts of the stellar 
mass bin $\log_{10}(M_{\star} [\Msun])=11.00-11.25$ and the distributions
predicted from our best-fit 
phase-space density model.

\subsection{Mass profile}

\begin{figure}
\begin{center}
    \leavevmode
    \epsfxsize=8cm
    \epsfbox[50 50 580 420]{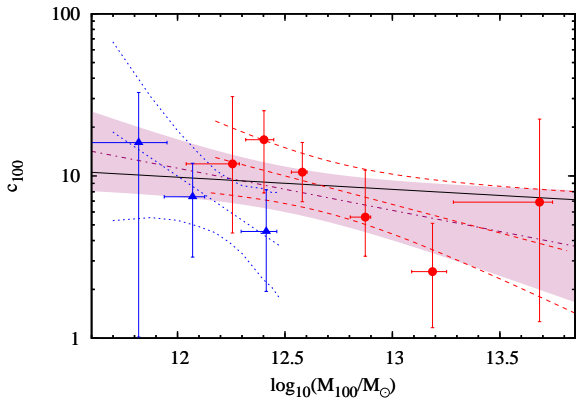}
\end{center}
\caption{Concentration-mass relation for galaxy-size haloes. The points with the error bars show constraints on the halo mass 
and the concentration parameter of the DM density profile inferred from the kinematics of the satellite galaxies 
(red circles for red hosts and blue triangles for blue hosts). The purple dashed-dotted line 
is the best-fit power-law fit and the shaded region is the $1\sigma$
confidence area calculated by bootstrapping from MCMC models. The
  red (dashed) and blue (dotted) line shows the best-fit profile ($\pm1\sigma$) with the
  data comprising red and blue hosts. The black solid line is the median
concentration-mass relation from cosmological simulations of a standard
$\Lambda$CDM model (the Bolshoi Simulation, \citealp{KTP11}).}
\label{con-mv}
\end{figure}

\begin{figure}
\begin{center}
    \leavevmode
    \epsfxsize=8cm
    \epsfbox[50 50 580 420]{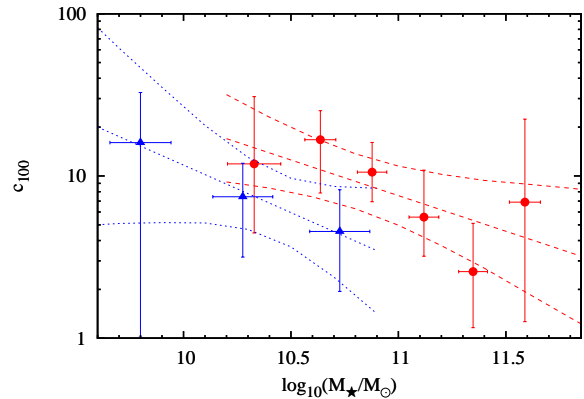}
\end{center}
\caption{Same as Fig.~\ref{con-mv}, but as a function of the stellar mass.}
\label{con-ms}
\end{figure}


Fig.~\ref{con-mv} shows the comparison between our `observational' constraints on the concentration-mass relation from satellite 
kinematics (red and blue points) and its predictions from cosmological
simulations  (black solid line). For the latter, we plotted 
the concentration-mass profile from the Bolshoi Simulation -- a high resolution simulation of a standard $\Lambda$CDM cosmological 
model with the most updated cosmological parameters \citep{KTP11}. The concentration parameters inferred from the satellite 
velocities are fairly consistent with the profile from cosmological
simulations. However, our constraints are not tight enough to determine robustly the slope of the mass-concentration relation. 
Power-law fits to the data of both types of the host galaxies yields the slope $-0.26\pm0.23$ (purple solid line in Fig.~\ref{con-mv}), where the error is calculated by bootstrapping MCMC models. This is consistent with 
the predicted value of $-0.075$ \citep{KTP11} as well as with a flat profile. The best-fit normalisation at $M_{100}=3\times10^{12}\Msun$ is $8.4\pm2.3$, in excellent agreement with the concentration $c_{100}=9.1$ 
from the simulations of the current $\Lambda$CDM cosmological model \citep{KTP11}.

Our concentration-mass relations for red and blue hosts are consistent with a flat profile. Fig.~\ref{con-mv} shows 
the best-fit power-law profiles obtained in the same way as above, but independently for the data of red and blue hosts. The best-fit slopes and normalisations at the median mass are: $-0.35\pm0.29$ and $c(5\times10^{12})=8.5\pm3.6$ 
(red hosts), 
$-0.91\pm0.95$ and $c(10^{12}\Msun)=10.0_{-4.8}^{+9.0}$ (blue hosts). 
Comparing the profiles at 
$M_{100}=2.5\times10^{12}\Msun$, which is a common mass scale of red and blue hosts in the sample, we find that the 
concentration parameter for blue galaxies is smaller than in the red: $c=4.4_{-2.0}^{+3.5}$ (blue hosts), 
$c=10.9_{-3.5}^{+5.2}$ (red hosts). This finding is significant at the $2\sigma$ level and is inconsistent with 
a standard CDM scenario of halo formation which does not predict any
dichotomy of DM concentration at fixed halo mass.

Fig.~\ref{con-ms} shows the concentration parameter as a function of the stellar mass. Fitting power-law 
profiles yields the slopes $-0.44\pm-.47$ ($-0.59\pm 0.7$) and the normalisations $c(10^{11}\Msun)=7.6_{-2.3}^{+3.4}$ ($c(2\times10^{12}\Msun)=7.7_{-3.5}^{+6.4}$) for the blue (red) hosts. Concentration parameter of the blue hosts tend 
to be smaller by factor of $\approx2$ than the red with the same stellar mass. Both $c-M_{\star}$ relations have 
comparable slopes. The stellar masses of the red hosts are typically larger by $0.7$ dex then those of the blue with 
the same concentration of dark matter.

\begin{figure}
\begin{center}
    \leavevmode
    \epsfxsize=8cm
    \epsfbox[50 50 580 420]{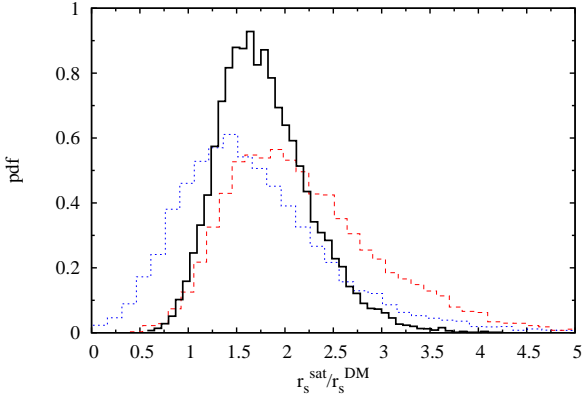}
\end{center}
\caption{Probability distribution for $r_{\rm s}^{\rm sat}/r_{\rm s}^{\rm DM}$ combined from all bins of the stellar mass, 
where $r_{\rm s}^{\rm sat}$ and $r_{\rm s}^{\rm DM}$ are the scale radii of the satellite number density profile and dark 
matter density profile. Red, blue and black colours (dashed, dotted and solid lines, respectively) correspond to the red and blue host galaxies, and both 
types of the central galaxies.
}
\label{ampl-prob}
\end{figure}

Fig.~\ref{ampl-prob} shows the probability distribution for $r_{\rm s}^{\rm sat}/r_{\rm s}^{\rm DM}$ combined from all bins of the stellar mass. 
The typical scale radius of the satellite number density profiles around red host galaxies is larger by factor of $1.8_{-0.6}^{+0.9}$ 
than that corresponding to DM ($r_{\rm s}^{\rm sat}/r_{\rm s}^{\rm DM}>1.0$ at the $97$ per cent confidence level). This finding 
confirms previous studies based on photometric data from the SDSS and showing that the spatial distribution of the satellite 
around red galaxies is more extended than of DM and is well-fitted by the NFW profile with the concentration parameter 
typically $2$ times smaller than that expected for DM \citep{GCEF12}. Constraints on the ratio of the satellite-to-dark-matter 
scale radius for blue galaxies do not point to any bias between the spatial distribution of DM and the satellites, i.e. 
$r_{\rm s}^{\rm sat}/r_{\rm s}^{\rm DM}=1.4_{-0.6}^{+0.8}$. Combining the results from both types of central galaxies yields the host-independent 
ratio $r_{\rm s}^{\rm sat}/r_{\rm s}^{\rm DM}=1.6_{-0.3}^{+0.5}$ ($r_{\rm s}^{\rm sat}/r_{\rm s}^{\rm DM}>1.0$ at the $97$ per cent confidence level). 
This estimate is fully consistent with the bias between the concentration parameters of the subhalo number density profile and dark 
matter density profile measured in cosmological simulations, $r_{\rm s}^{\rm sat}/r_{\rm s}^{\rm DM}=1.5$ from the Millennium Simulation 
\citep{Sales+07} and $r_{\rm s}^{\rm sat}/r_{\rm s}^{\rm DM}=1.3$ from the
Bolshoi Simulation \citep{KTP11}.

\subsection{Halo-stellar mass relation}

\begin{figure}
\begin{center}
    \leavevmode
    \epsfxsize=8cm
    \epsfbox[50 50 580 420]{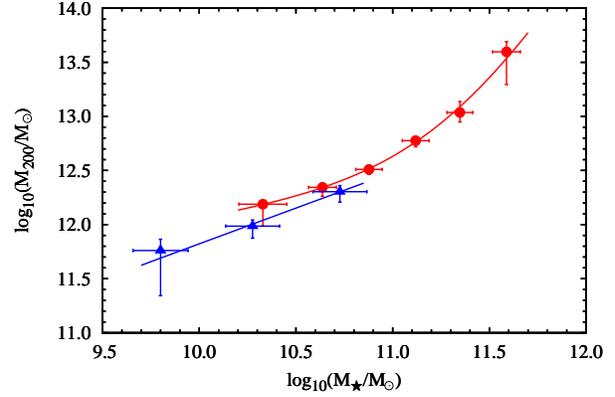}
\end{center}
\caption{Halo versus stellar mass for red and blue galaxies (red circles and blue triangles, 
respectively). The error bars indicate the range containing $68$ per cent of the marginal probability distribution for the halo 
mass and the scatter of the stellar masses in bins. Solid lines are the best fit power-law profile for blue galaxies and 
the best fit double-power-law profile given by (\ref{ms-mv-fit}) for red
galaxies.}
\label{ms-mv}
\end{figure}

Fig.~\ref{ms-mv} shows the halo mass as a function of the stellar mass of both types of host galaxies. The halo mass is 
measured with accuracy up to $0.05$ dex for red galaxies with stellar masses $\log_{10}(M_{\star}/\Msun)=10.75-11.25$ for 
which number of the satellites per stellar mass bin reaches maximum. Our constraints on the halo-stellar mass relation 
reveal a change of the slope between low and high stellar masses, with the transition mass 
$\log_{10}(M_{\star}/\Msun)\approx {11}$. Following \citet{Dutton+10}, we find that a reasonable fit to the data is achieved using 
the following function
\begin{equation}
M_{200}=M_{{\rm h}\,0}\left({M_{\star}\over M_{\star\,0}}\right)^{a}
\left[\frac{1}{2}+\frac{1}{2}\,\left({M_{\star}\over M_{\star\,0}}\right)^{\gamma}\right]^{(b-a)/\gamma},
\label{ms-mv-fit}
\end{equation}
where $\alpha$ and $\beta$ are the logarithmic slopes at small and large
stellar masses, respectively, 
$M_{\star\,0}$ and $M_{{\rm h}\,0}$ are the stellar and halo mass at the transition point, and $\gamma$ is a parameter 
controlling the sharpness of the transition. Fitting this function to the data of red galaxies yields $a=0.29$, 
$b=2.91$, $\gamma=1.24$, $\log_{10}(M_{\star\,0}/\Msun)=11.3$ and
$\log_{10}(M_{h\,0}/\Msun)=13.1$. 
Our constraints on the halo-to-stellar mass relation for blue galaxies cover approximately one order of magnitude in the stellar mass 
and are consistent with a power-law with logarithmic slope $0.66\pm 0.07$ and normalisation 
$\log_{10}(M_{200}/\Msun)=12.0$ at the stellar mass
$\log_{10}(M_{\star}/\Msun)=10.3$. These best-fit relations
of the halo-stellar mass relation are also shown in Fig.~\ref{ms-mv}.

\begin{figure}
\begin{center}
    \leavevmode
    \epsfxsize=8cm
    \epsfbox[50 50 580 420]{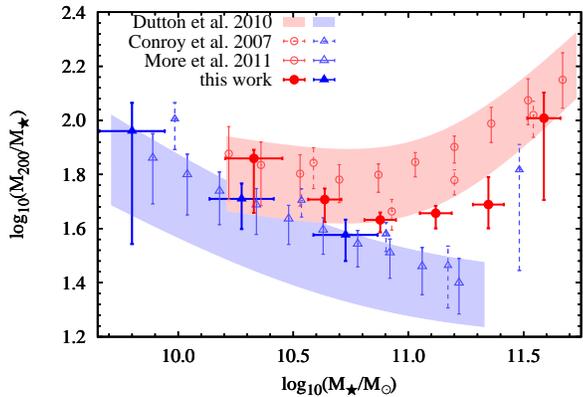}
\end{center}
\caption{Halo-to-stellar mass ratio vs. stellar mass (same symbols and colours as in Fig.~\ref{ms-mv}) 
compared to other measurements selected from 
the literature. Shaded stripes (left: blue hosts, right: red hosts) represent a compilation of 
results comprising constraints based on the satellite kinematics and weak lensing \citep{Dutton+10}. 
Empty symbols show constraints based on the satellite kinematics obtained by \citet{Conroy+07_dyn} 
and \citet{More+11}, respectively (blue hosts with dashed error bars). All error bars and the widths of 
the stripes are $68$ per cent or $1\sigma$.}
\label{comparison}
\end{figure}

Fig.~\ref{comparison} shows a comparison with other constraints on the halo-stellar mass relation selected 
from the literature. In particular, we refer to the existing constraints based on satellite kinematics 
\citep{Conroy+07_dyn,More+11} and the empirical halo-stellar mass relation obtained by \citet{Dutton+10} as 
a compilation of all available results based on both stellar kinematics \citep{Conroy+07_dyn,More+11} and 
weak lensing \citep{Mandelbaum+06_mhalo,MSH08,SMP10}. Stellar masses were converted to a common standard 
consistent with the \citet{Chabrier03} IMF, as described in \citet{Dutton+10}. We find that the halo-stellar mass relation 
from our analysis of red hosts is fairly consistent with other measurements. Compared to the results obtained 
by \citet{More+11}, it exhibits a slightly sharper transition between low and high stellar mass regime. 
This tension seems to be alleviated when comparing with results obtained by
\citet{Conroy+07_dyn}. But at the high end, our halo masses are typically 0.2 dex lower than found
in the literature \citep{Dutton+10}. 

Our constraints for blue hosts agree with those from \citet{More+11}. On the other hand, halo masses 
appear to be offset by $0.1$ dex with respect to the compiled profile obtained by \citet{Dutton+10}. This 
trend occurs for the measurements from \citet{Conroy+07_dyn} and \citet{More+11}, suggesting that 
the weak lensing technique leads to preferentially lower halo masses in late-type galaxies than 
the stellar kinematics.

\begin{figure}
\begin{center}
    \leavevmode
    \epsfxsize=8cm
   \epsfbox{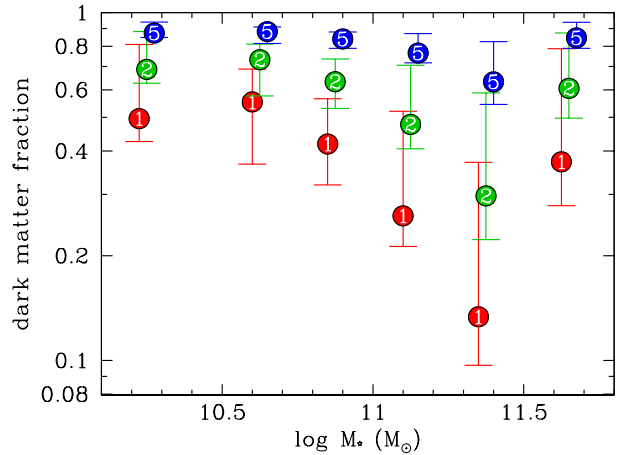}
\end{center}
\caption{Dark matter fraction versus stellar mass for red host galaxies, at
  1 (red), 2 (green), and 5 (blue) $R_e$. The abscissae for the different
  values of $r/R_e$ are slightly shifted for clarity. The error bars are 16th and 84th
  percentiles from the
  marginalised 
  distributions.}
\label{fracdm}
\end{figure}

Figure~\ref{fracdm} shows the DM fraction at 1, 2, and $5\,R_e$ as a function
of stellar mass, for red host galaxies (for details of the calculation of the stellar 
mass distribution, see Summary and discussion)
While the DM within the virial
radius is least important for $10.75 < \log M_\star < 11$,
Figure~\ref{fracdm} indicates that the
DM fraction at several effective radii is minimised at a larger mass
interval: $11.25 < \log M_\star
< 11.5$, which happens to be the one for which DM halos have the lowest DM
concentrations (Fig.~\ref{con-mv}).

\subsection{Anisotropy of the satellite orbits}

Constraints on the anisotropy parameter obtained in individual bins of the stellar mass are not tight enough to draw any 
solid conclusion.  For example, working in separate bins of host stellar
mass,
we cannot  differentiate between an isotropic velocity distribution ($\beta=0$) and the typical 
anisotropy profile found in simulated DM haloes where $\beta$ increases with radius from $0.1$ in the halo centre 
to $0.3-0.5$ at the virial radius \citep{WLGM05,AG08,CPKM08}. Since there is no theoretical hint that the anisotropy 
may depend on the halo mass and results obtained in different stellar mass bins do not reveal any trend with the stellar 
mass, it is advisable to combine constraints from all bins into one. 

\begin{figure}
\begin{center}
    \leavevmode
    \epsfxsize=8cm
    \epsfbox[50 50 580 580]{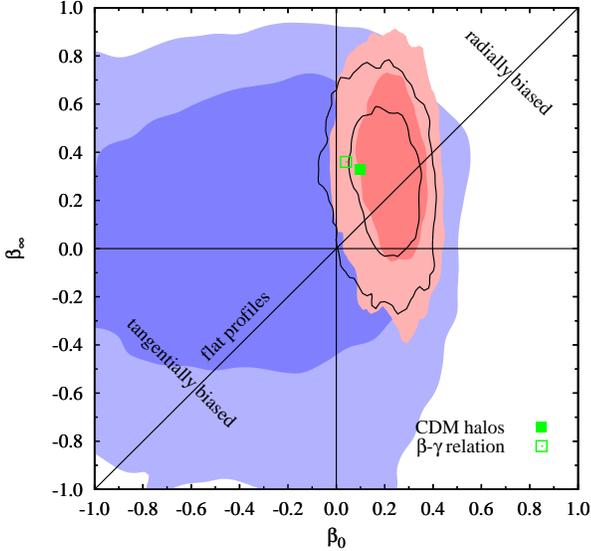}
\end{center}
\caption{Constraints on the asymptotic values of the anisotropy profile parameter $\beta(r)$, $\beta_{0}=\beta(r\ll r_{\rm s}^{\rm sat})$ and $\beta_{\infty}=\beta(r\gg r_{\rm s}^{\rm sat})$. 
The contours are the confidence regions containing $68$ and $95$ per cent of the probability 
combining results from the stellar mass bins. Red, blue and both types of host galaxies are shown 
as red shaded regions, (wide) blue shaded regions, and black contours, respectively. Green filled and 
empty squares indicate the typical 
$\beta(r)$ profile of $\Lambda$CDM halos \citep[based on][]{AG08,Wojtak+08} and the 
universal relation between the anisotropy and DM density slope
\citep{HM06,HJS10} applied to our best fit NFW model. 
}
\label{beta-prob}
\end{figure}

Fig.~\ref{beta-prob} shows contours of the resulting joint 
probability distribution of two parameters determining the anisotropy profile,
$\beta_{0}=\beta(r\ll r_{\rm s}^{\rm sat})$ and 
$\beta_{\infty}=\beta(r\gg r_{\rm s}^{\rm sat})$. When we combine  all
host galaxy mass bins, we find that the satellite 
orbits are mildly radially anisotropic with $\beta_{0}=0.2\pm0.1$ and $\beta_{\infty}=0.3\pm0.2$. 
These constraints on $\beta(r)$ profile are not significantly different from typical 
anisotropy profiles of $\Lambda$CDM halos \citep[red square][]{AG08,Wojtak+08}, neither from the universal 
relations between the anisotropy and the logarithmic slope of an underlying DM density profile 
\citep{HM06,HJS10}. The shift between the marginal probability distributions for $\beta_{\infty}$ and $\beta_{0}$ may indicate a tendency for the anisotropy to increase with radius. 
This effect, however, is of a marginal statistical significance and the data still permit a flat 
anisotropy profile.

Our measurement of the anisotropy parameter relies mostly on the data for red galaxies (compare the red and black probability 
distributions in Fig.~\ref{beta-prob}). Combining results for red galaxies yields $\beta_{0}=0.26_{-0.10}^{+0.08}$ and 
$\beta_{\infty}=0.38^{+0.26}_{-0.27}$ which are statistically consistent with the constraints obtained for both types of central 
galaxies. Measurement of the orbital anisotropy around blue galaxies is much weaker due to significantly smaller number 
of the satellites and it does not allow to differentiate between an isotropic velocity distribution and the anisotropy profile 
motivated by cosmological simulations (see the blue probability distribution
in Fig.~\ref{beta-prob}). The analysis of the combined 
probability distribution results in $\beta_{0}=-0.2_{-0.6}^{+0.4}$ and $\beta_{\infty}=0.2_{-0.5}^{+0.3}$.

\section{Summary and discussion}

We made use of satellite kinematics compiled from the redshift catalog of the
SDSS to measure the physical properties of dark 
matter haloes and the orbital anisotropy of the satellites orbiting fairly isolated galaxies. Our data analysis was carried out in 
the framework of the projected phase-space density based on an anisotropic model of the distribution function for equilibrated 
spherical systems \citep{Wojtak+08}. This approach avoids arbitrary binning the data. It
furthermore  allows us to break the mass-anisotropy
degeneracy \citep{WLMG09}.

We found that the relation between halo masses and stellar mass of galaxies
matched, to first order, the previous measurements through satellite
kinematics \citep{Conroy+07_dyn,More+11} and weak lensing (compiled together by \citealp{Dutton+10}).
In particular, we confirm that the halo masses of red hosts are significantly
greater than those of blue hosts of same stellar mass. However, at the high
end of our red hosts, our halo masses are typically 0.2 dex lower than found
in the literature \citep{Dutton+10}. This might be the consequence of a stricter
isolation criterion applied to our host galaxies, thus avoiding better host galaxies
at the centres of groups.

The concentration parameter of DM density profiles has a typical value of
$\approx 9$, in full consistency with the results from cosmological 
simulations \citep{KTP11}. Our constraints on the concentration-mass relation are not tight enough to determine its slope over 
the range of galactic halo masses. However, a more robust measurement of the
$c-M_{100}$ slope may be achieved by means of 
complementing these results by similar constraints at higher halo masses
available in the literature. Combining the normalisation 
of the $c-M_{100}$ resulting from satellite kinematics with that obtained by \citet{WL10} from galaxy kinematics in clusters 
($c_{100}=6.8\pm0.7$ at $M_{100}=4.9\times 10^{14}\Msun$) yields the slope $-0.05\pm0.04$, in fair agreement with $-0.075$ 
from cosmological simulations \citep{KTP11}.

We found that red hosts have significantly more concentrated DM halos than blue hosts of the same stellar
or halo mass. 
Na\"{\i}vely, one would conclude that
the halos of red hosts assembled earlier than those of blue hosts, since halo
concentrations tend to be greater at higher redshift \citep{ZMJB03}. However,
if red galaxies are built by mergers of blue galaxies, the halo mass of the
merger remnant  will be close to the sum of the original ones, while the
concentration will remain the same if the merger keeps the DM density profile
self-similar, as found in binary major mergers of NFW models \citep{KZK06}. 
Given the negative slope of the $c-M$ relation, this would
explain why red galaxies have higher halo concentration than blue galaxies of
same halo mass. However the offset seen in Figure~\ref{con-mv} between the red and blue host galaxy $c-M$
relation is roughly a factor of 3 at given halo mass, so one would have to
conclude that red galaxies are the products of more than a single major
merger of blue galaxies. 

Mass modelling of elliptical galaxies using stellar kinematics is usually
limited to 3--4 effective radii, beyond which the signal-to-noise ratio of
spectra are too low to properly infer line-of-sight velocity dispersions. Our
analysis of the satellite kinematics allows us to probe down to $5\,R_e$. We
find (top broken line of Fig.~\ref{fracdm}) that, at this radius, the DM
component always dominates, but less so for 
stellar masses $11.25 < \log M_\star < 11.5$. If we extrapolate our analysis to
lower radii, we obtain the same trend with stellar mass (bottom two broken
lines of Fig.~\ref{fracdm}). In particular, at $2\,R_e$, the DM component
should dominate at all stellar masses except for $11.25 < \log M_\star <
11.5$. It will be worth confronting this (extrapolated) prediction with forthcoming
observational studies of elliptical galaxy internal kinematics out to
$2\,R_e$ in a large range of stellar masses.

Although our analysis relies on a specific parameterisation of the density
profile, the existence of a well-constrained dark 
matter scale radius $r_{\rm s}^{\rm DM}<r_{100}$ suffices to conclude that the observed satellite kinematics is fully compatible 
with the NFW density profile of DM and can hardly be reconciled with an isothermal sphere model suggested in 
several studies based on lensing analyses of massive elliptical galaxies \citep{Koopmans+06,Koopmans+09,Gavazzi+07}. An isothermal density profile of DM 
would noticeably affect the measurement of the concentration parameter
resulting in $r_{\rm s}^{\rm DM}\approx r_{100}$. 
Note that the weak lensing measurements of \citet{Leauthaud+10} are
consistent with both power-laws and NFW DM profiles (see their Fig.~4).
So the apparent flattening of the density profile at large radii implied by
isothermal profiles 
should probably be attributed to a projection effect of the local dense environment rather than to DM haloes.
This 
finding confirms and complements a number of tests showing consistency of satellite kinematics with the NFW profile 
of DM density \citep{Prada+03,KP09}.

In our analysis, by adopting a single NFW model for the host mass distribution, we have neglected the contribution of the stellar component. 
Since elliptical galaxies are known to be dominated by their stellar component within the effective radius \citep{ML05a,Humphrey+06}, one may 
worry that our DM concentrations will be overestimated, even though we only considered satellites further than $5\,R_e$ from the host galaxy. For 
example, the total density profile of a two-component NFW+S\'ersic model is very close to a singular isothermal in the range $0.1 \to 1\,R_e$ 
(see upper left panel of Fig.~4 of \citealp{ML05b}), thus explaining the
isothermal profile found in this fairly low range of radii by 
\citet{Koopmans+06}.  Since the mass distribution returned from our model is most sensitive to the line-of-sight velocity dispersion profile, 
$\sigma_{\rm los}(R)$, we asked ourselves how much lower would the DM concentration be if we incorporated a \cite{Sersic68} model to the mass 
distribution.

We performed this test for our red host galaxies. For each of our bins of stellar mass, we determined the median $r$-band absolute magnitude. 
We then obtained from Table~3 of \cite{Simard+11} the effective radii, $R_e$, and S\'ersic indices, $n$, for these absolute magnitudes, after restricting 
the SDSS sample of \citeauthor{Simard+11} to the Red Sequence (using the same cut as we did in our mass analysis) and our adopted redshift range. We then 
computed $\sigma_{\rm los}(R)$ for the satellite population with a two-component NFW+S\'ersic mass model, adopting the central stellar mass of our mass 
bin with the values of $R_e$ and $n$ that we obtained above, the satellite scale radius that 
we previously measured (derived from Table~\ref{tab-res}), adopting $\beta(r) = 0.2$ for the 
satellites (consistent with Fig.~6), as well as the DM normalisation derived from our 1-component mass model (Table~\ref{tab-res}). 
The DM concentration, $c_{100}$, 
is a free parameter. We iterated on $c_{100}$ until our profile of $\sigma_{\rm los}(R)$ matched the profile expected for a 1-component NFW model. 
For each of the six mass bins, we were then able to match $\sigma_{\rm
  los}(R)$ to better than 1 per cent typically (5 per cent in the worst case
of stellar mass and radius) between $5\,R_e$ and the 
virial radius $r_{100}$. In the end, we found that the 
concentration of the DM component was 7 to 20 per cent lower than in the 1-component model. This suggests that our derived values of $c_{100}$ are overestimated by 7 
to 20 per cent. Note that if we allowed for adiabatic contraction (e.g., \citealp{Gnedin+11}), the DM concentration would be greater, hence our overestimate would 
be lower. This simple analysis suggests that our choice of a single NFW model for the mass distribution of isolated galaxies beyond $5\,R_e$ is reasonable, 
as the predicted line-of-sight velocity dispersion profiles match that of more realistic two component models and our concentration is typically overestimated 
by only 10 per cent.

Satellite kinematics reveals a bias between the spatial distributions of DM and satellite galaxies. The scale 
radius of the satellite number density profile is typically larger by factor of $1.6$ than of DM. This 
finding is consistent with the estimate of a counterpart bias between DM particles and subhaloes found 
in cosmological simulations \citep[see e.g.][]{Sales+07,KTP11}. We did not find any statistically significant difference 
between the bias of red and blue galaxies.

The orbital anisotropy of satellite galaxies exhibits a mild excess of radial orbits with typical anisotropy parameter 
$\beta=0.2\pm0.1$ in the inner regions and $\beta=0.3\pm0.2$ in the outer
regions of their hosts. These constraints on the inner and outer asymptotic
values of the anisotropy profile are statistically consistent with the values
found in the halos in $\Lambda$CDM cosmological simulations.
The difference between inner and outer anisotropies is 
too weak to reveal any statistically 
significant trend of the anisotropy with radius. 

Due to the substantial difference between the numbers of satellites around red and blue hosts, 
our constraints on the orbital anisotropy come
principally from the satellites around red hosts. 
This radial anisotropy around red (giant elliptical) galaxies has been
predicted from hydrodynamical simulations of binary
mergers \citep{Dekel+05}. Some giant elliptical galaxies shows signs of such radial outer
anisotropy \citep{Das+08, deLorenzi+08}, while others do not \citep{Napolitano+11}.
The velocity distribution of the satellites orbiting blue galaxies is less
well constrained, and is consistent with an isotropic model, contrary to the
case for the satellites orbiting red hosts.

In general, fitting an anisotropic ($\beta\ne 0$) model of galaxy kinematics leads to different mass profiles 
than when isotropic velocities ($\beta=0$) are forced. For example, the same velocity dispersion profile may be  equally well fit
by a steep mass profile with isotropic orbits or a shallow mass profile with radially biased orbits \citep[see][]{Merritt87}. 
In order to assess the impact of the anisotropy on our results, we reanalysed the data assuming an isotropic model of 
the phase-space density ($\beta_{0}=\beta_{\infty}=0$). We found that the new constraint on the virial masses and the scale 
radii $r_{\rm s}^{\rm sat}$ remained the same within the errors. On the other
hand, concentration parameters of dark matter profiles and $r_{\rm s}^{\rm
  sat}/r_{\rm s}^{\rm DM}$ ratios for red hosts tended to be larger by typically $17$ 
percent, which is comparable to the error on the normalisation of the mass-concentration relation. This steepening 
of DM mass profiles is the only effect of ignoring the anisotropy of satellites orbits.

\section*{Acknowledgments}
The authors thank an anonymous referee for the comments and help in improving the manuscript. 
RW warmly thanks Aaron Dutton for sharing his data compilation with constraints on the halo-stellar mass relation. 
The Dark Cosmology Centre is funded by the Danish National Research Foundation. RW is grateful for the 
hospitality of Institut d'Astrophysique de Paris where part of this work was done. RW thanks Steen Hansen 
for fruitful discussions and Anna Gallazzi for her useful advice about stellar masses of galaxies from 
SDSS. The computations were performed on the facilities provided by the Danish Center for Scientific 
Computing.

\bibliography{master}

\end{document}